\documentstyle[12pt]{article}
{\baselineskip 12pt}
\begin{document}
\title{{The operator form of the effective potential governing the time evolution in $n$-
dimensional subspace of states}}
\author{\\J. Piskorski$^{\ast}$\\ \hfill \\
Pedagogical University, Institute of Physics, \\
Plac Slowianski 6, 65-069 Zielona Gora, Poland.}
\maketitle

\maketitle
\nopagebreak
\begin{abstract}
\noindent This paper presents the operator form of the effective potential $V$ governing the time 
evolution in  $2$ and $3$ and $n$ dimensional subspace of states. The general formula for the $n$ 
dimensional case is considered the starting point for the calculation of the explicit formulae 
for $2$ and $3$ dimensional degenerate and non-degenerate cases. We relate the $2$ and $3$ 
dimensional cases to some physical systems which are currently investigated.
\end{abstract}
$^{\ast}$e--mail: jaro@magda.iz.wsp.zgora.pl; \\

\section[1]{Introduction}
The time evolution of physical systems in the Hilbert space is described by the Shr\"odinger 
equation: \begin{equation}
i \frac{\partial}{\partial t} |{\psi};t> = H 
|{\psi};t>,  \label{Sch}
\end{equation}
where
$\hbar = c = 1$.\\
If we choose the initial conditions:
\begin{equation}
|\psi;t=0>\equiv|\psi>,
\label{initialcond}
\end{equation}
then the time evolution is described by a unitary operator $U(t)|\psi>=|\psi;t>$ 
($|\psi>,|\psi;t>\in \cal{H}$, $U(t)=e^{-itH}$). 

Vector $|\psi;t>\in \cal{H}$ carries complete information about
the physical system considered. In particular, the properties  of the system which are described 
by vectors belonging to a closed subspace $\cal{H}_{||}$, of 
$\cal H$ can be extracted from 
$|\psi;t>$ .  In such a case it is sufficient to know  the component
$| \psi ;t>_{||} \in {\cal H}_{||}$ of $| \psi ; t>$. The subspace 
$\cal{H}_{||}$ is defined by a projector $P$: $\cal{H}_{||} = P {\cal H}$,
which simply means that $| \psi ;t>_{||} = P| \psi ;t>$.

Alternatively, the same result can be obtained by studying the time evolution not in the total 
space of states $\cal{H}$ but in 
a closed subspace 
$\cal{H}_{||}$. 
In this way the 
total state space is split into two orthogonal subspaces $\cal{H}_{\parallel}$ and 
$\cal{H}_{\perp}=\cal{H}\ominus\cal{H}_{\parallel}$, and 
the  Shr\"odinger equation can be replaced by equations describing each of the  subspaces 
respectively. The equation for $\cal{H}_{\parallel}$ has the following form \cite{1}---\cite{3}:
\begin{equation}
\left(i\frac{\partial}{\partial t}-PHP\right)|\psi;t>_{||}=| \chi ;t>-
i\int_{0}^{\infty}K(t-\tau)|\psi;\tau>_{||}d\tau, \label{evolutparal}
\end{equation}
\begin{equation}
Q=I-P,
\end{equation}
\begin{equation}
K(t)=\Theta(t)PHQe^{-iQHQ}QHP,
\end{equation}
\begin{equation}
|\chi;t>=PHQe^{-iQHQ}|\psi>_{\perp},
\end{equation}
where
\[
\Theta(t)=\left\{
\begin{array}{lll}
1&{\rm for}&t\geq0\\
0&{\rm for}&t<0 \end{array}
\right.
.\]
Of course $K(t)\neq 0$ only if $[P,H]\neq0$. Condition (\ref{initialcond}) can now be rewritten 
as
\begin{equation}
|\psi;t=0>_{||}\equiv|\psi>_{||},   |\psi;t=0>_{\perp}\equiv|\psi>_{\perp},
\label{initialcond2}
\end{equation}
where $|\psi>_{\perp}\equiv Q|\psi>.$\\
If we now assume that at the initial moment no states from $\cal{H}_{\perp}$ are occupied, 
$|\psi>_{\perp}=0$, ( that is $|\chi;t>\equiv0$, $|\psi>\equiv\psi>_{||})$ and define the 
evolution operator for the subspace $\cal{H}_{||}$:
\begin{equation}
|\psi;t>_{||}\equiv P|\psi;t>\equiv PU(t)|\psi>\equiv PU(t)P|\psi>_{||},
\end{equation}
so
\begin{equation}
U_{||}(t)|\psi>_{||}\stackrel{\rm def}{=}PU(t)P|\psi>_{||},
\label{u}
\end{equation}
we can transform (\ref{evolutparal}) into
\begin{equation}
\left(i\frac{\partial}{\partial t}-PHP\right)U_{||}|\psi>_{||}=-
i\int_{0}^{\infty}K(t-\tau)U_{||}(\tau)|\psi>_{||}d\tau.
\label{evolutparal2}
\end{equation}
An equivalent differential form of (\ref{evolutparal2}) has been found by Kr\'olikowski and 
Rzewuski \cite{1,2}:
\begin{equation}
\left(i\frac{\partial}{\partial t}-H_{||}(t)\right)U_{||}(t)|\psi>_{||}=0,
\verb+   + t\geq0, \verb+   + U_{||}(0)=P,
\label{krolik}
\end{equation}
where the $H_{||}(t)$ denotes the effective Hamiltonian: \begin{equation}H_{||}(t)\equiv 
PHP+V_{||}(t)\label{hparalel}.\end{equation}\\
For every effective Hamiltonian $H_{\parallel}$  governing  the time
evolution in
 ${\cal H}_{\parallel}$ $\equiv P {\cal H}$, which in
general  can  depend on  time  $t$  \cite{1} --- \cite{3}, the
following identity holds \cite{4}-\cite{7}:

\begin{equation}
H_{\parallel}(t)  \equiv
i \frac{{\partial}U_{\parallel}(t)}{\partial t}
[U_{\parallel}(t)]^{-1} P,
\label{l5}
\end{equation}
where $[U_{\parallel}(t)]^{-1} $, is defined as follows
\begin{equation}
[U_{\parallel}(t)]^{-1} U_{\parallel}(t) =
U_{\parallel}(t) [U_{\parallel}(t)]^{-1}\equiv P.
\label{l6}
\end{equation}
In the nontrivial case
\begin{equation}
[P,H] \neq 0,
\label{l7}
\end{equation}
from (\ref{l5}), using  (\ref{hparalel}) and (\ref{u}) we find

\begin{eqnarray}
H_{\parallel}(t) & \equiv &  PHU(t)P[U_{\parallel}(t)]^{-1} P
\label{l7a1}  \\
& \stackrel{\rm def}{=} & PHP + V_{\parallel}(t).  \label{l7b}
\end{eqnarray}
and thus
\[V_{||}(t)\equiv PHQ U(t) [ U_{\parallel}(t)]^{-1}P \label{l7a}.\]
Assumption (\ref{l7}) means  that  transitions
of states from ${\cal H}_{\parallel}$ into ${\cal H}_{\perp}$  and
from ${\cal H}_{\perp}$ into ${\cal H}_{\parallel}$, i.e., the decay
and regeneration processes, are allowed. Thus \cite{4} ---
\cite{6},
\begin{equation}
H_{\parallel}(0) \equiv PHP, \; \; V_{\parallel}(0) = 0,
\; \; V_{\parallel} (t \rightarrow0) \simeq -itPHQHP,
\label{l8}
\end{equation}
so,   in   general   $H_{\parallel}(0)    \neq$
$H_{\parallel}(t \gg t_{0}=0)$ \cite{4} --- \cite{7} and
$V_{\parallel}(t \neq 0) \neq V_{\parallel}^{+}(t \neq 0)$,
$H_{\parallel}(t \neq 0) \neq H_{\parallel}^{+}(t \neq 0)$.
According to the ideas of the standard scattering theory, it  can be stated  that  operator 
$H_{\parallel}(t \rightarrow \infty)$
$\equiv H_{\parallel}(\infty)\stackrel{\rm def}{=}H_{||}$ describes the bounded or
quasistationary states  of  the  subsystem  considered  (and  in
this  sense  it is similar to e.g. the LOY--effective Hamiltonian \cite{8}).

From (\ref{evolutparal2}) and (\ref{krolik}),(\ref{hparalel}) it follows that the action of 
$V_{||}(t)$ on $U_{||}(t)$ has the following form:
\begin{equation}
V_{||}(t)U_{||}(t)=-i\int^{\infty}_{0}K(t-\tau)U_{||}(\tau)d\tau.
\label{vonu}
\end{equation}
The approximate form of $V_{||}$ can be obtained from (\ref{evolutparal2}) and (\ref{vonu}) with 
the use of the retarded solution of: 
\begin{equation}
\left(i\frac{\partial}{\partial t}-PHP\right)G(t)= P\delta(t),
\end{equation}
where $G(t)$ is the retarded Green operator:
\begin{equation}
G\equiv G(t)=-i\Theta(t)e^{-itPHP}P.
\end{equation}
Then, using the iteration procedure for the equation (\ref{evolutparal2}) for $U_{||}$ \cite{2}, 
\cite{5,6,7} we get:
\begin{equation}
U_{||}=U^{0}_{||}(t)+\sum_{n=1}^{\infty}(-i)^{n}\underbrace{L\circ L\circ L\circ... \circ L}_{n 
\verb+ + times}\circ U^{0}_{||}(t).
\label{convol}
\end{equation}
$U^{0}_{||}(t)$ is the solution of the following "free" equation \cite{5,6,7}:
\begin{equation}
\left(i\frac{\partial}{\partial t}-PHP\right)U_{||}^{0}(t)= 0,\verb+   + U_{||}^{0}(0)=P.
\end{equation} 
$\circ$ stands for the convolution 
\[f\circ g(t)=\int^{\infty}_{0}f(t-\tau)g(\tau),\]
and 
\[L\equiv L(t)=G\circ K(t).\]
Equations (\ref{vonu}) and (\ref{convol}) yield:
\begin{equation}
V_{||}(t)U_{||}(t)=-iK\circ U^{0}_{||}(t)+\sum_{n=1}^{\infty}(-i)^{n}K\circ L\circ L\circ... 
\circ L\circ U^{0}_{||}(t).
\label{fullvonu}
\end{equation}
If $||L(t)||<1$ then the series (\ref{fullvonu}) is convergent. It is worth noticing that, unlike 
in the standard perturbation series, it is not necessary for the perturbation $H^{1}$ to be small 
in relation to $H^{0}$ (the full Hamiltonian $H=H^{0}+H^{1}$) if $||L(t)||<1$. This is considered 
one of the advantages of this approach over the standard ones as it can describe both week and 
strong interactions \cite{6}.\\
If for every $t\geq0$ $||L(t)||<1$ then to the lowest order of $L(t)$, $V_{||}(t)$ is expressed 
by \cite{6}:
\begin{equation}
V_{||}(t)\simeq V^{1}_{||}(t)=-i\int^{\infty}_{0}K(t-\tau)e^{i(t-\tau)PHP}Pd\tau.
\label{valpar}
\end{equation}
This formula was used to compute $V_{||}(t)$ for 
one--dimensional subspace ${\cal H}_{\parallel}$  and to 
find the matrix elements of $V_{||}(t)$ acting in a two--state subspace 
${\cal H}_{\parallel}$ in \cite{7}.  In some problems it is more useful and more convenient to 
use the operator form of $V_{\parallel}(t)$ rather than the 
 the matrix elements of $V_{\parallel}(t)$ only. Searching for
the global transformation properties of $V_{\parallel}(t)$
under some operators expressing symetries of the system 
is as an example of such problems. 

Result (\ref{valpar}) will be the starting point for the following 
considerations concerning the explicit operator form of $V_{||}(t)$ in 
$n$, $2$ and $3$ dimensional cases.

\section[2]{Effective potential $V_{\parallel}$ in n-dimensional\\ subspace of states.}
Let us consider a general case of effective potential $V_{\parallel,n}$, acting in an $n$-
dimensional subspace of states. Formally, the equation corresponding to Eq.(\ref{hparalel}) has 
the following form:
\begin{equation}
H_{\parallel,n}(t) \stackrel{\rm def}{=}  PHP + V_{\parallel,n}(t).
\label{H-par-defn} 
\end{equation}
The projector $P$ is defined in the following way:
\begin{equation}
P = \sum_{j=1}^{n} |{\bf e}_{j}><{\bf e}_{j}| \equiv I_{\parallel},
\label{P-nd}
\end{equation}
where $I_{\parallel}$ is the unit operator in $\cal{H_{||}}$, $\verb+{+|{\rm \bf 
e}_{j}>\verb+}+_{j\in {\cal A}}$ and $\verb+{+|{\rm \bf 
e}_{j}>\verb+}+_{j=1,2,...n}\subset\verb+{+|{\rm \bf e}_{j}>\verb+}+_{j\in {\cal A}}$ are 
complete sets of orthonormal vectors $<{\rm \bf e}_{j}|{\rm \bf e}_{k}>=\delta_{jk}$ in $\cal{H}$ 
and $\cal{{H}_{\parallel}}\subset\cal{H}$ respectively. Consequently, if the state space for the 
problem is $\cal{H}$ then $\cal{{H}_{\parallel}}=P\cal{H}$ and $P$ is the unity in 
$\cal{H}_{\parallel}$, $P=I_{\parallel}$ \cite{7}.\\
The subspace $\cal{H}_{\parallel}$ can also be spanned by the eigenvectors of the hermitian 
matrix $PHP$:
\begin{equation}
PHP |{\bf \lambda}_{j}> = {\bf{\lambda}}_{j} |{\bf \lambda}_{j}>, \; \;
\scriptscriptstyle{(j=1,2,...,n)}.
\label{l-j}
\end{equation}

Using $|\lambda_{j}>$ we define projectors $P_{j}$ \cite{9} where  for simplicity the non-
degenerate case of $\lambda_{j}$ is assumed. :
\begin{equation}
P_{j} \stackrel{\rm def}{=} 
\frac{1}{<{\lambda}_{j} |{\lambda}_{j} >}
|{\lambda}_{j}><{\lambda}_{j}|\label{P-j}, \verb+   + \scriptscriptstyle{(j=1,2,3)}.
\end{equation}
Of course these projectors fulfill the following completness condition:
\begin{equation}
\sum_{j=1}^{n}P_{j}=P.\label{completness}
\end{equation}
The operator $PHP$ can now be written as follows:
\begin{equation}
PHP = \sum_{j=1}^{n} {\lambda}_{j} P_{j}, 
\label{PHP-sp}
\end{equation}
and following:
\begin{equation}
P e^{\textstyle \pm itPHP} = P \sum_{j=1}^{n} 
e^{\pm it {\lambda}_{j} }P_{j}.
\label{ex-PHP-sp}
\end{equation}
This result can be directly applied to equation (\ref{valpar})
\begin{equation}
V_{\parallel,n}(t)
=  -i \sum_{j=1}^{n} \int_{0}^{t}
PHQ e^{\textstyle -i (t - \tau ) (QHQ - {\lambda}_{j})} QHP
\,   d \tau  \, P_{j} . \label{V-3d2}
\end{equation}
The integration can be easily performed, with the result:
\begin{equation}
V_{\parallel,n}(t)= -i
\sum_{j=1}^{n} \Big\{
PHQ \frac{e^{-it(QHQ - \lambda_{j})}-1}{QHQ - \lambda_{j}}QHP \Big\}P_{j}=
\sum_{j=1}^{n}\Xi(\lambda_{j},t)P_{j},
\label{earlyV}
\end{equation}
where
\begin{equation}
\Xi(\lambda,t)\stackrel{\rm def} {=} PHQ \frac{e^{-it(QHQ - \lambda)}-1}{QHQ -\lambda}QHP.
\end{equation}
Knowing that \cite{10}
\begin{equation}
\lim_{t\rightarrow\infty}\Xi(\lambda,t)=PHQ \frac{1}{QHQ - \lambda+i0}QHP,
\end{equation}
and defining
\[\Sigma(\lambda)\stackrel{\rm def}{=}PHQ \frac{1}{QHQ - \lambda+i0}QHP\],
we finally get:
\begin{equation}
V_{\parallel,n}\stackrel{\rm def}{=}\lim_{t\rightarrow \infty}V_{\parallel,n}(t)=  -
i\lim_{t\rightarrow \infty }
\sum_{j=1}^{n}\Xi(\lambda_{j},t)P_{j}=-i\sum_{j=1}^{n}\Sigma(\lambda_{j})P_{j}.
\label{wzornav}
\end{equation}
\section[3]{$V_{\parallel}$ in a two dimensional subspace.}
In this section we find the explicit formula for $V_{\parallel}$ in a two-dimensional subspace of 
states using the framework presented above.\\ In this case $PHP$, being a $[2\times2]$ hermitian 
matrix, has the following form:
\begin{equation}
PHP=\left[\begin{array}{rr}
H_{11}&H_{12}\\
H_{21}&H_{22}\end{array}\right],
\end{equation}
\[H_{ij}=H^{\ast}_{ji},\]
where $H_{j,k}=<{\rm \bf e}_{j}|H|{\rm \bf e}_{k}>$.\\
The eigenvalues of $PHP$ are easy to calculate:
\begin{equation}
PHP|\lambda>=\left[\begin{array}{rr}
H_{11}&H_{12}\\
H_{21}&H_{22}\end{array}\right]
\left(\begin{array}{r}
\alpha_{1}\\
\alpha_{2}\end{array}\right)=\lambda\left(\begin{array}{r}
\alpha_{1}\\
\alpha_{2}\end{array}\right),
\end{equation}
\begin{equation}
\lambda^{1,2}=\frac{1}{2}(H_{11}+H_{22})\pm \sqrt{|H_{12}|^{2}+\frac{1}{4}(H_{11}-H_{22})},
\end{equation}
and if we adopt the symbols used in \cite{4}-\cite{7}:
\begin{equation}
\lambda^{1,2}\stackrel{\rm def}{=}H_{0} \pm \kappa,\end{equation}
where
\begin{equation}
H_{0}=\frac{1}{2}(H_{11}+H_{22}),\verb+   + \kappa=\sqrt{|H_{12}|^{2}+\frac{1}{4}(H_{11}-
H_{22})}.
\end{equation}
Following, the eigenvector $|\lambda^{1}>$ can be chosen as follows:
\begin{equation}
|\lambda^{1}>=\left(\begin{array}{c}
\frac{H_{12}}{H_{0} + \kappa - H_{11}}\\
1\end{array}\right),
\end{equation}
and the projector $P_{1}$:
\begin{equation}
P_{1}= 
\frac{1}{<{\lambda}_{1} |{\lambda}_{1} >}
|{\lambda}_{1}><{\lambda}_{1}|=
\end{equation}
\[=\frac{1}{\frac{|H_{12}|^{2}}{(H_{0} + \kappa - H_{11})^{2}}+1}\left(\begin{array}{c}
\frac{H_{12}}{H_{0} + \kappa - H_{11}}\\
1\end{array}\right)\cdot\left(\frac{H_{21}}{H_{0} + \kappa - H_{11}},1\right),
\]
so, explicitly
\begin{equation}
\large{P_{1}=\left[\begin{array}{cc}
\frac{(H_{0} + \kappa - H_{11})\cdot|H_{12}|^{2}}{|H_{12}|^{2}+(H_{0} + \kappa - H_{11})^{2}}&
\frac{(H_{0} + \kappa - H_{11})\cdot H_{12}}{|H_{12}|^{2}+(H_{0} + \kappa - H_{11})^{2}}\\
\frac{(H_{0} + \kappa - H_{11})\cdot H_{21}}{|H_{12}|^{2}+(H_{0} + \kappa - H_{11})^{2}}&
\frac{(H_{0} + \kappa - H_{11})^{2}}{|H_{12}|^{2}+(H_{0} + \kappa - 
H_{11})^{2}}\end{array}\right]}.
\end{equation}
For clarity let us define:
\[P_{j}\stackrel{\rm def}{=}\left[\begin{array}{rr}
p^{j}_{11}&p^{j}_{12}\\
p^{j}_{21}&p^{j}_{22}\end{array}\right] \verb+     +\scriptscriptstyle{(j=1,2)}.\]
Both $P_{1}$ and $PHP$ can be represented by Pauli matrices:
\[P_{1}=p^{1}_{0}\sigma_{0}+p^{1}_{x}\sigma_{x}+p^{1}_{y}\sigma_{y}+p^{1}_{z}
\sigma_{z},\]
\[PHP=H_{0}\sigma_{0}+H_{x}\sigma_{x}+H_{y}\sigma_{y}+H_{z}
\sigma_{z},\]
 and the calculation of the coefficients $p^{j}$ yields:
\begin{equation}
\large{\begin{array}{l}
p_{0}=\frac{1}{2}(p_{11}+p_{22})=\frac{1}{2},\\
p_{x}=\frac{1}{2}(p_{12}+p_{21})=\frac{H_{x}}{2\kappa},\\
p_{y}=\frac{1}{2}i(p_{12}-p_{21})=\frac{H_{y}}{2\kappa},\\
p_{z}=\frac{1}{2}(p_{11}-p_{22})=\frac{H_{z}}{2\kappa}.\end{array}}
\end{equation}
We can see from the above that $p_{\nu}, (\nu=0,x,y,z)$ can be expressed by $H_{\nu}, 
(\nu=0,x,y,z)$, so finally we get the following expression for $P_{1}$
\begin{equation}
P_{1}=
\large{\frac{1}{2}\left( \left(1-\frac{H_{0}}{\kappa}\right)\sigma_{0}+\frac{1}{\kappa}PHP
\right)}.
\end{equation}
Keeping in mind the fact that in $\cal{H}_{\parallel}$ we have $\sigma_{0}=I_{\parallel}=P$, we 
obtain:
\begin{equation}
P_{1}=\large{\frac{1}{2}\left( \left(1-\frac{H_{0}}{\kappa}\right)P+\frac{1}{\kappa}PHP\right)},
\end{equation}
and after performing the same calculation for $P_{2}$:
\begin{equation}
P_{2}=\large{\frac{1}{2}\left( \left(1+\frac{H_{0}}{\kappa}\right)P-\frac{1}{\kappa}PHP\right)}.
\end{equation}
It is easy to verify that the completness condition (\ref{completness}) is fulfilled:
\[P_{1}+P_{2}=P.\]
If we now come back to Eq.(\ref{earlyV}) and use the results obtained in this section, the 
effective potential $V_{\parallel}$ will have the following form:
\begin{eqnarray}
V_{\parallel}(t)& = & - \frac{1}{2} \Xi (H_{0} + 
\kappa,t ) \Big[ (1 - \frac{H_{0}}{\kappa} )P +
\frac{1}{\kappa} PHP \Big] \nonumber \\
& \; & - \frac{1}{2} \Xi (H_{0} - 
\kappa,t ) \Big[ (1 + \frac{H_{0}}{\kappa} )P -
\frac{1}{\kappa} PHP \Big] . \label{V-imp2} 
\end{eqnarray}
Matrix elements of this $V_{\parallel}(t)$ are exactly the same as those obtained in \cite{7}.

As noted in \cite{7} this result is significant. For example in the case of neutral $K$ mesons 
the assumption of $CPT$ invariance and $CP$ noninvariance in the quantun theory, that is 
$[CPT,H]=0$ and $[CP,H]\neq 0$, yields:
\begin{equation}
h_{11}-h_{22}\neq0 ,
\label{neq}
\end{equation}
where $h_{ij}=<{\rm \bf e}_{i}|H_{||}|{\rm \bf e}_{j}>$ are the matrix elements of $H_{||}\equiv 
PHP+V_{||}$, $V_{||}\stackrel{\rm def}{=}V_{||}(\infty)$, which runs counter to the usual 
assumption. More remarks on this problem can be found in the conclusions.

The case of both eigenvalues of $PHP$ equal can easily be obtained from the general case 
described above. The assumption of both eigenvalues equal for a hermitian $[2\times2]$ matrix 
yields $H_{11}=H_{22}$ and $H_{12}=H_{21}=0$.  It is easy to verify that 
$\lambda^{1}=\lambda^{2}\Longleftrightarrow\kappa=0$. Then:
\begin{equation}
\lambda^{1}=\lambda^{2}=H_{0}
\end{equation}
\[PHP=H_{0}P\]
and
\begin{equation}
Pe^{itPHP}\equiv e^{itH_{0}}P
\end{equation}
Thus, from equations (\ref{valpar}) and (\ref{V-3d2}):
\begin{equation}
V_{||}(t)\simeq V^{1}_{||}(t)=-\Xi(H_{0},t)P
\label{vtwodegener}
\end{equation} 
Furthermore, if apart from assuming the degenerate case of $PHP$ we take $t\rightarrow\infty$ we 
will get the same result as obtained from the Wigner-Weisskopf approximation by e.g. Lee, Oehme 
and Yang \cite{8}. It is interesting to notice that in this case $h_{11}=h_{22}$ (where 
$h_{jj}=<j|H_{||}|j>$) with $[CPT,H]=0$, whereas in the case of $\lambda^{1}\neq\lambda^{2}$ 
under the same conditions we have (\ref{neq}).

\section[4]{$V_{\parallel}$ in a three dimensional subspace.}
This section describes the explicit formula for $V_{\parallel}$ in a three dimensional subspace 
of states in a very similar way as it was done for the two dimensional case. 

In this case the $PHP$ matrix is a $[3\times 3]$ matrix, for example
\begin{equation}
PHP=\left[\begin{array}{rrr}
H_{11}&H_{12}&H_{13}\\
H_{21}&H_{22}&H_{23}\\
H_{31}&H_{32}&H_{33}\end{array}\right],
\end{equation}
\[H_{ij}=H^{\ast}_{ji},\]
and has the following characteristic equation:
\begin{equation}
\lambda^{3}+A\lambda^{2}+B\lambda+C=0,\\
\label{character}
\end{equation}
\[
\begin{array}{l}
A=-(H_{11}+H_{22}+H_{33}),\\
B=H_{11}H_{22}+H_{11}H_{33}+H_{22}H_{33}-|H_{13}|^{2}-|H_{23}|^{2}-|H_{13}|^{2},\\
C=-(H_{11}H_{22}H_{33}+2Re(H_{12}H_{23}H_{31})-H_{11}|H_{23}|^{2}
-H_{22}|H_{13}|^{2}-H_{33}|H_{12}|^{2}).\\
\end{array}
\]

It is easy to notice that $A,B,C\in\Re$ so, given the fact that $PHP$ is a hermitian matrix, 
equation (\ref{character}) is a third order equation with real coefficients and real solutions. 
To find the solutions we will use the Cardano formulae. Bearing in mind that the solutions are 
real we get the following three cases $\lambda_{1}\neq\lambda_{2}\neq\lambda_{3}$, 
$\lambda_{1}=\lambda_{2}=\lambda\neq\lambda_{3}$ and 
$\lambda_{1}=\lambda_{2}=\lambda_{3}=\lambda$:
Let us find the eigenvectors, projectors and the quasipotential for each of the above cases.
\subsection{$\lambda_{1}\neq\lambda_{2}\neq\lambda_{3}$.} In this case the three solutions of the 
characteristic equation (\ref{character}) are given by the following formulae:
\begin{equation}
\begin{array}{l}
\lambda_{1}=-2(\frac{A^{2}-3B}{3})^{\frac{1}{2}}\cos\frac{1}{3}\alpha-\frac{A}{3},\\
\lambda_{2}=-2(\frac{A^{2}-3B}{3})^{\frac{1}{2}}\cos\frac{1}{3}(\alpha+2\pi)-\frac{A}{3},\\
\lambda_{3}=-2(\frac{A^{2}-3B}{3})^{\frac{1}{2}}\cos\frac{1}{3}(\alpha+4\pi)-\frac{A}{3},\\
\end{array}
\label{solutionthree}
\end{equation} 
where $\cos\alpha=\frac{\frac{3}{2}(\frac{2A^{3}}{27}-\frac{B}{3}+C)}{\frac{2}{3}(\frac{A^{2}-
3B}{3})^{\frac{3}{2}}}.$\\
The following basis of orthogonal eigenvectors can be chosen:
\begin{equation}
|\lambda_{j}>=\left(\begin{array}{c}
H_{13}(H_{22}-\lambda_{j})-H_{23}H_{12}\\
H_{23}(H_{11}-\lambda_{j})-H_{13}H_{21}\\
|H_{12}|^{2}-(H_{11}-\lambda_{j})(H_{22}-\lambda_{j})
\end{array}\right),
\label{eigenthree}
\end{equation}
where $j=1,2,3$.\\
Using these eigenvectors we create projectors $P$ in the way given in Sec.2.:
\begin{equation}
P_{j}=\frac{1}{<\lambda_{j}|\lambda_{j}>}|\lambda_{j}>
<\lambda_{j}|=
\label{projthree}
\end{equation}
\[\begin{array}{l}
\{ |H_{13}(H_{22}-\lambda_{j})-H_{23}H_{12}|^{2}+\\
|H_{23}(H_{11}-\lambda_{j})-H_{13}H_{21}|^{2}+\\
\left[|H_{12}|^{2}-(H_{11}-\lambda_{j})(H_{22}-\lambda_{j})\right]^{2}\}^{-1}\times
\end{array}
\]
\[
\times\left[\begin{array}{rrr}
p^{j}_{11}&p^{j}_{12}&p^{j}_{13}\\
p^{j}_{21}&p^{j}_{22}&p^{j}_{23}\\
p^{j}_{31}&p^{j}_{32}&p^{j}_{33}\end{array}\right], \verb+   + \scriptscriptstyle{(j=1,2,3)},
\]
where 
\[\begin{array}{l}
p^{j}_{11}=|H_{13}(H_{22}-\lambda_{j})-H_{23}H_{12}|^{2},\\
p^{j}_{12}=(H_{13}(H_{22}-\lambda_{j})-H_{23}H_{12})(H_{32}(H_{11}-\lambda_{j})-H_{31}H_{12}),\\
p^{j}_{13}=(H_{13}(H_{22}-\lambda_{j})-H_{23}H_{12})(|H_{12}|^{2}-(H_{11}-\lambda_{j})(H_{22}-
\lambda_{j})),\\
p^{j}_{21}=(H_{23}(H_{11}-\lambda_{j})-H_{13}H_{21})(H_{31}(H_{22}-\lambda_{j})-H_{32}H_{21}),\\
p^{j}_{22}=|H_{23}(H_{11}-\lambda_{j})-H_{13}H_{21}|^{2},\\
p^{j}_{23}=(H_{23}(H_{11}-\lambda_{j})-H_{13}H_{21})(|H_{12}|^{2}-(H_{11}-\lambda_{j})(H_{22}-
\lambda_{j})),\\
p^{j}_{31}=(|H_{12}|^{2}-(H_{11}-\lambda_{j})(H_{22}-\lambda_{j})(H_{31}(H_{22}-\lambda_{j})-
H_{32}H_{21}),\\

p^{j}_{32}=(|H_{12}|^{2}-(H_{11}-\lambda_{j})(H_{22}-\lambda_{j})(H_{32}(H_{11}-\lambda_{j})-
H_{31}H_{12}),\\

p^{j}_{33}=(|H_{12}|^{2}-(H_{11}-\lambda_{j})(H_{22}-\lambda_{j}))^{2},
\end{array},
\]
(where j=1,2,3).\\
And consequently the quasipotential 
\begin{equation}
V_{\parallel,3}(t)= -i
\sum_{j=1}^{3}\Xi(\lambda_{j},t)P_{j},
\label{vthree}
\end{equation}

\subsection{$\lambda_{1}=\lambda_{2}=\lambda\neq\lambda_{3}$.} In this case we have the following 
expressions for the solutions of the characteristic equation (\ref{character}):
\begin{equation}
\begin{array}{l}
\lambda=(\frac{2A^{3}}{27}-\frac{B}{3}+C)^{\frac{1}{3}}-\frac{A}{3},\\
\lambda_{3}=-2(\frac{2A^{3}}{27}-\frac{B}{3}+C)^{\frac{1}{3}}-\frac{A}{3}.
\end{array}
\label{solutiontwo}
\end{equation}
In this case to define one of the projectors, say $P_{3}$  we can use the result presented above, 
so the projector will be given by formula (\ref{projthree}).\\
We do not actually need to know the remaining two projectors explicitly as
\begin{equation}
V_{\parallel,3}(t)= -i
\Xi(\lambda,t)(P_{1}+P_{2})+\Xi(\lambda_{3},t)P_{3}
\label{vtwo}
\end{equation}
and $P_{1}+P_{2}=P-P_{3}$, $P$ is the unity in the considered space so:
\begin{equation}
V_{\parallel,3}(t)= -i
\Xi(\lambda,t)(P-P_{3})+\Xi(\lambda_{3},t)P_{3}
\label{vtwoa}
\end{equation}
\subsection{$\lambda_{1}=\lambda_{2}=\lambda_{3}=\lambda$.} This case is the simplest one, and 
the solutions are:
\begin{equation}
\lambda=-\frac{A}{3}=H_{11}=H_{22}=H_{33}
\end{equation}
In this case $PHP$ is a diagonal matrix in any basis. In fact, this is true for any $n$-
dimensional hermitian matrix with all eigenvalues equal, so we get a form of quasipotential which 
is identical to the two dimensional degenerate case (\ref{vtwodegener}).
\begin{equation}
V_{\parallel}=-\Xi(\lambda,t)P
\end{equation}
Again, if apart from assuming the three-fold degenerate case of $PHP$ we take 
$t\rightarrow\infty$ we will get a result which is analogous to the one obtained from the Wigner-
Weisskopf approximation by e.g. Lee, Oehme and Yang \cite{8}.
\section{Equation for $\rho$ matrix in ${\cal H}_{||}$ . }

This section contains one possible application of the result obtained above, which is the 
equation for the density matrix $\rho$ in ${\cal H}_{||}$.\\

Very often systems of the type described in Section 1. are considered as open systems
interacting with  an  unknown  rest,  i.e.,  with  the  reservoir
\cite{11,12}.  Then,  for  the  description  of the time  evolution in
 subspace ${\cal H}_{\parallel}$, instead  of  the state   vector
$|\psi ;t>_{\parallel}$ solving equations (\ref{evolutparal}), (\ref{krolik}), density
matrix $\rho$ is used. 
The $\rho$--matrix  in quantum mechanics fulfills the following  equation:

\begin{equation}
\frac{\partial}{\partial t} \rho =
i [ \rho ,H] ,
\label{b1}
\end{equation}
where  $H$  is  the  total  Hamiltonian  of   the   system   under
consideration acting in the Hilbert state space $\cal H$. $H$  and
$\rho$  are  hermitian. \\
 The consideration of such systems sometimes begins 
with  a   phenomenological   Hamiltonian
$H_{eff} \equiv  H_{\parallel}$, acting in an $n$-dimensional subspace ${\cal H}_{\parallel}$.
Such Hamiltonians are of the LOY  type or the type used in the master  equation  approaches 
\cite{11,12}. 
These Hamiltonians are not hermitian, therefore
the  time evolution of the reduced $\rho$--matrix,  i.e.,
${\rho}_{\parallel}$ (where ${\rho}_{\parallel}$ denotes the part of
$\rho$--matrix acting in  ${\cal H}_{\parallel}$ ),  is  given  by
\cite{11}
\begin{equation}
\frac{\partial}{\partial t}  {\rho}_{\parallel} =
- i \Big( H_{\parallel} {\rho}_{\parallel}
- {\rho}_{\parallel} H_{\parallel}^{+} \Big) ,
\label{b2}
\end{equation}
where
\begin{equation}
{\rho}_{\parallel}(t) \equiv
\left(
\begin{array}{cccccc}
{\rho}_{11}(t) & {\rho}_{12}(t) & \ldots &\rho_{1n}(t) & 0 & \ldots  \\
{\rho}_{12}^{\ast}(t) & {\rho}_{22}(t) & \ldots &\rho_{2n}(t) & 0 & \ldots \\
\ldots & \ldots & \ldots & \ldots & 0 & \ldots \\ 
\rho_{1n}^{\ast}(t) &\rho_{2n}^{\ast}(t)&\ldots &\rho_{nn}(t)& 0 &\ldots \\
0 & 0 & 0 & 0 & 0 & \ldots  \\
\ldots & \ldots & \ldots & \ldots & \ldots &\ldots 
\end{array}
\right)
\label{b3}
\end{equation}
At this point one remark concerning the above should be made: all properties of
${\rho}_{\parallel}(t \; > \; 0)$ solving this evolution equation
are determined  by  the
form and properties  of  $H_{\parallel}$ so  for  the  same  initial
conditions ${\rho}_{0}$ but different $H_{\parallel}$ a different ${\rho}_{\parallel}(t)$ can be 
obtained.

Let us notice that the solution of Eq.(\ref{b1}) has the  following
form
\begin{equation}
\rho (t) \equiv
U(t) {\rho}_{0} U^{+}(t),
\label{b4}
\end{equation}
where ${\rho}_{0} \equiv \rho (0)$ and $U(t)$ is  the  total
unitary evolution operator for the system considered. From this we
conclude that  the  exact  reduced  $\rho$--matrix  for  a  given
complete and closed subspace ${\cal H}_{\parallel}$ of the
total state space $\cal H$ is
\begin{equation}
{\rho}_{\parallel} (t) \equiv P \rho (t) P,
\label{b5}
\end{equation}
If the subsystem described by ${\rho}_{\parallel}(t)$ is an open
system, i.e., if transitions from subspace ${\cal H}_{\parallel}$
into ${\cal H} \ominus  {\cal H}_{\parallel}$  (and  vice  versa)
occur, then $P$ cannot commute  with  the  total  Hamiltonian  $H$.

Now, in order to describe an $n$ state system of the considered type, ${\rho}_{0}$ of the form 
(\ref{b3}) and a projector  defining  the
subspace  of the   form   (\ref{P-nd}),   or
another unitary one equivalent to it, should be chosen. It is easy  to  verify  that
for this $P$ we have
\begin{equation}
{\rho}_{0} \equiv P {\rho}_{0} P ,
\label{b6}
\end{equation}
so, in this case (see (\ref{b5}) and (\ref{b4}))
\begin{equation}
{\rho}_{\parallel}(t) \equiv P \rho (t) P
\equiv P U(t) P {\rho}_{0} P U^{+}(t) P .
\label{b7}
\end{equation}
Using the identity (\ref{u}) we have
\begin{equation}
{\rho}_{\parallel} (t) \equiv
U_{\parallel} (t) {\rho}_{0} U_{\parallel}^{+} (t) .
\label{b8}
\end{equation}
It can be easily verified that  ${\rho}_{\parallel}(t)$   fulfills   the
following equation,
\begin{equation}
i \frac{\partial}{\partial t} {\rho}_{\parallel} (t)
= \Big( i \frac{\partial U_{\parallel}(t)}{\partial t} \Big)
{\rho}_{\parallel} (t) + {\rho}_{\parallel} (t)
\Big( i \frac{\partial U_{\parallel}^{+}(t)}{\partial t}
\Big) ,
\label{b9}
\end{equation}
or, equivalently
\begin{equation}
i \frac{\partial}{\partial t} {\rho}_{\parallel} (t)
\equiv H_{\parallel} (t) \; {\rho}_{\parallel}(t)
- {\rho}_{\parallel}(t) \; H_{\parallel}^{+} (t) ,
\label{b10}
\end{equation}
(where $H_{\parallel}(t)$ is given by the identity (\ref{l5})),
which is analogous to (\ref{b2}).
\section[7]{Conclusions}
This paper deals with the operator form of the effective potential governing the time evolution 
in $n$-dimensional subspace of states. The general expression for such an effective potential has 
been found in Section 2. Sections 3. and 4. dealt with the explicit form of such an operator for 
$2$ and $3$ dimensional cases. In Section 5. an application of the formalism developed in the 
previous sections to the density matrix has been suggested.

The approach presented in this paper can be considered a natural extention of the Wigner-
Weisskopf approach to the single line width to more level subsystems which interact with the rest 
of the physical system. It has been shown that in the case of $n$ level systems the WW approach 
may only be suitable if the $PHP$ is $n$ - fold degenerate, which of course is not always the 
case. 

The physical problem which is currently investigated with the use of si\-mi\-lar methods is the 
neutral kaon complex and the possible violation of the $CPT$ symmetry. This problem is obviously 
a $2$ dimensional problem and can be researched with the use of the formalism developed in 
Section 3. The standard approach to the problem developed in \cite{8} uses the WW approximation 
to describe the time evolution of the $K_{0}, \overline{K_{0}}$ complex and proves to be quite a 
successful approximation of the physical reality. As noted at the end of Section 3., one of the 
conclusions which can be drawn here is that $h_{11}=h_{22}$. This can be measured, and the 
parameter $\delta_{CPT}\sim h_{11}-h_{22}$ is used in tests of $CPT$ conservation. However, if we 
want to retain the geometry of the problem (i.e. we do not want to reduce the problem to a one 
dimensional problem by assuming $PHP$ degeneration) we will find that $\delta_{CPT}\neq0$ even 
under  $CPT$ conserved. For a more extensive discussion of this problem see \cite{4,5,7}.

The three dimensional case has not as yet been applied to describe an actual physical system and 
the possibilities of doing so will be investigated in future papers. One possiblility is to use 
the density matrix approach which has been proposed in Section 5., to the description of multi-
level atomic transitions. Experiments designed to demonstrate the Quantum Zeno effect provide an 
example of such multi-level systems. For example Cook suggested an experiment which should 
demonstrate this effect on an induced transition in a single, trapped ion \cite{13}. This 
experiment assumes the ion to have a $3$ -- level structure, and to describe it the density 
matrix approach is usually used (see for example \cite{14}). This gives us a possiblity to use 
the results obtained in Section 4.1  ($\lambda_{1}\neq\lambda_{2}\neq\lambda_{3}$) and Section 5. 
to construct a suitable equation for the reduced three dimensional density matrix. This, however, 
is beyond the scope of this paper and, as noted earlier, will be researched in future papers.

\end{document}